\documentclass[%
 reprint,
 amsmath,amssymb,
 aps,
prb
]{revtex4-1}

\usepackage{graphicx}
\usepackage{dcolumn}
\usepackage{bm}

\begin{document}


\title{Recurrent network classifier for ultrafast skyrmion dynamics  }

\author{ A. Y. Deviatov, I. A. Iakovlev, V. V. Mazurenko}
%
\affiliation{
Theoretical Physics and Applied Mathematics Department, Ural Federal University, Mira Street 19, Ekaterinburg 620002, Russia\\
}
\date{\today}

\begin{abstract}
By using the supervised learning we train a recurrent neural network to recognize and classify ultrafast magnetization processes realized in two-dimensional nanosystems with Dzyaloshinskii-Moriya interaction. Our focus is on the different types of skyrmion dynamics driven by ultrafast magnetic pulses. Each process is represented as a sequence of the sorted magnetization vectors inputted into the network. The trained network can perform an accurate classification of the skyrmionic processes at zero temperature in wide ranges of parameters that are the magnetic pulse width and damping factor. The network performance is also demonstrated on different types of unseen data including finite temperature processes. Our approach can be easily adapted for creating an autonomous control system on skyrmion dynamics for experiments or device operations.  
\end{abstract}

\pacs{Valid PACS appear here}
\maketitle

\section{Introduction}
Understanding of relaxation processes in magnetic materials and systems due to the ultrafast pulses realized with different experimental techniques plays a crucial role for creating next-generation magnetic storage, spintronics and quantum technologies \cite{trotzky, ultrafast1,ultrafast2,ultrafast3}. Nowadays pico- and femtosecond magnetization dynamics of a system is accessible with X-ray free electron lasers \cite{XFEL} and Lorentz microscopy \cite{Lorentz}. At these time scales it becomes possible to trace the behaviour of magnetic systems frame by frame or spectrum by spectrum. The amount of such a data grows rapidly and one of the important problems is classification of the dynamical processes. The main specific here is that each experimental frame is a part of a particular excitation process and can not be analyzed separately, which makes this problem to be similar to classification of video content \cite{video1}. 

In material science a special focus is on the magnetic skyrmions that are topologically protected multi-spiral magnetic excitations formed in two- and three-dimensional materials. Each skyrmion is a complex object with non-collinear ordering of the atomic magnetic moments stabilized by the Dzyaloshinskii-Moriya interaction. In the static case the magnetic state recognition and classification problems can be effectively solved neither with standard methods of machine learning such as support vector machine, k-means and others or with neural network approach \cite{Iakovlev1,Iakovlev2}. The latter is very useful for analysis of the transitional areas between different phases, for instance, between skyrmion and spin spiral or between skyrmion and ferromagnetic states. Nevertheless, the elementary functions (erase and write) with skyrmions for next-generation electronics are entirely related to dynamics of skyrmions in external electric or magnetic fields \cite{Nagaosa}. A practical implementation of these elementary functions requires an autonomous control on the system's dynamics, which motivates us to develop a neural network approach for recognition and classification of dynamical skyrmionic processes.

\begin{figure}[t]
\center 
\includegraphics[width=\columnwidth]{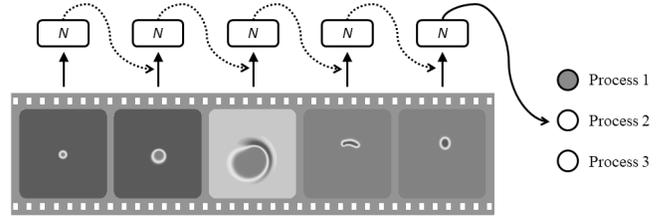}
\caption{Illustration of idea of the ultrafast skyrmionic process recognition. Magnetization dynamics is used frame by frame as an input for recurrent neural network providing the process classification.} \label{Opening}
\end{figure}

In our work we follow the idea ''classify a movie rather than snapshot'' \cite{Evert} and propose a recurrent neural network classifier for ultrafast skyrmion processes of picosecond scale in two-dimensional materials. A simplified scheme of our approach is presented in Fig.~\ref{Opening}. A stream of magnetic process frames is inputted into a recurrent neural network (RNN) frame by frame. We show that the process classification can be performed by tracing only the dynamics of the $z$ components of spins. The trained network provides robust results independent on the frame time frequency and pulse details during data collection. Our study paves the way to construct precise process diagrams for magnetic systems in wide range of external parameters.

\section{Model, Method and Network}
In this work we use the following spin Hamiltonian to describe two-dimensional materials hosting skyrmions
\begin{equation}\label{Ham}
\begin{split}
H= -\sum_{i\neq j}J_{ij}{\bf S}_i{\bf S}_j-\sum_{i\neq j}{\bf D}_{ij}[{\bf S}_i\times{\bf S}_{j}] -\sum_{i}K({\bf S}_i^z)^2.
\end{split}
\end{equation}
 Here $J_{ij}$ and ${\bf D}_{ij}$ are the isotropic exchange interaction and Dzyaloshinskii-Moriya vector, respectively. ${\bf S}_{i}$ is a unit vector along the direction of the $i$th spin, $K$ is the strength of the uniaxial anisotropy in the $z$ direction. We take into account the only interactions between the nearest neighbours. The summation for inter-spin couplings runs twice over every pair. The Hamiltonian is defined on the 100$\times$100 square lattice. In our simulations we used the following parameters: $J = 0.03676$ mRy, $|{\bf D}| = 0.008824$ mRy and $K = 0.00735$ mRy. Dzyaloshinskii-Moriya interaction vector for each pair of spins  is parallel to the corresponding inter-site radius vector. Such a Hamiltonian was previously used in Ref.~\onlinecite{Blugel} to simulate switching of magnetic skyrmions by picosecond magnetic field pulses via transient topological states.

Since we are interested in simulation of dynamical processes, to solve Eq.~\eqref{Ham} we used the Landau-Lifshitz-Gilbert (LLG) equation as it was implemented in the Uppsala Atomistic Spin Dynamics (UppASD) package \cite{UppASD1, UppASD2}

\begin{figure}[t]
\center 
\includegraphics[width=\columnwidth]{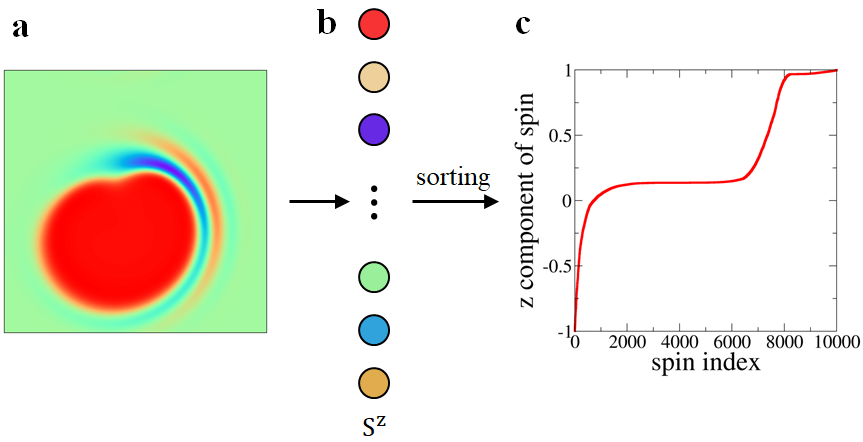} 
\caption{Schematic representation of used preprocessing procedure. Complete magnetic structure (a) of a  system is reduced to unsorted vector (b) containing $S^z$ components of spins. Red and blue circles correspond to the spins with z projections of 1 and -1, respectively. (c) Visualization of the sorted spin vector used as input to neural network. } \label{preproc}
\end{figure}

\begin{equation}\label{LLG}
\begin{split}
\frac{d\textbf{S}_i}{dt}= -\frac{\gamma}{1+\alpha^2}\textbf{S}_i\times[-\frac{\partial H}{\partial\textbf{S}_i}+b_i(t)]-\\ -\frac{\gamma}{|\textbf{S}_i|}\frac{\alpha}{1+\alpha^2}\textbf{S}_i\times(\textbf{S}_i\times[-\frac{\partial H}{\partial\textbf{S}_i}+b_i(t)]),
\end{split}
\end{equation}
where $\gamma$ is the gyromagnetic ratio, $\alpha$ is the damping parameter and $b_i(t)$ is a stochastic magnetic field with a Gaussian distribution arising from the thermal fluctuations. In principle, one can train a deep neural network to simulate the magnetization dynamics of the system, as it was done in Ref.~\onlinecite{Kovacs}.

To realize different processes related to skyrmion dynamics we used a time-dependent magnetic field pulse defined by a Gaussian distribution as it was proposed in Ref.~\onlinecite{Blugel}

\begin{equation}\label{Gauss}
\begin{split}
\textbf{B}_p(t)= B_0exp\left(-\frac{(t-t_p)^2}{2t_w^2}\right)\textbf{e}_B,
\end{split}
\end{equation}
where $B_0$ is the amplitude of the magnetic field, $t_w$ is the Gaussian width and $t_p$ is the time position of the pulse maximum. The real-space orientation of the magnetic pulse $\textbf{e}_B$ is described by the polar and azimuthal angles $\theta$ and $\varphi$. We used $B_0 = 2$ T and $\varphi = 0^{\circ}$. Other parameters were varied depending on the type of the simulation, as described below.

{\it Recurrent Neural Network}---
The magnetic configurations calculated with LLG approach at different times of the relaxation process are used as input for the recurrent neural network. We have found that it is of crucial importance to perform a data preprocessing that is ordering of the magnetization vector containing only z components of spins as shown in Fig.~\ref{preproc}. Namely, such a sorting provides a very accurate separation of the magnetic phases in the static case as discussed in Ref.~\onlinecite{Iakovlev2}. It also gives us opportunity to use a moderate number of the hidden layer neurons that is 512 for process classification. Without sorting one needs to substantially increase the number of the hidden layer neurons up to 5000 to get reasonable results.

To train the recurrent neural network we use magnetic processes generated with the $\theta = 50^\circ$ magnetic pulses. For the training set we generated 400 configurations for each of breathing, switching and collapse processes from the internal areas of the corresponding diagram. A complete description of the technical details concerning the neural network structure and training is presented in Appendices.

\begin{figure}[b]
\center 
\includegraphics[width=\columnwidth]{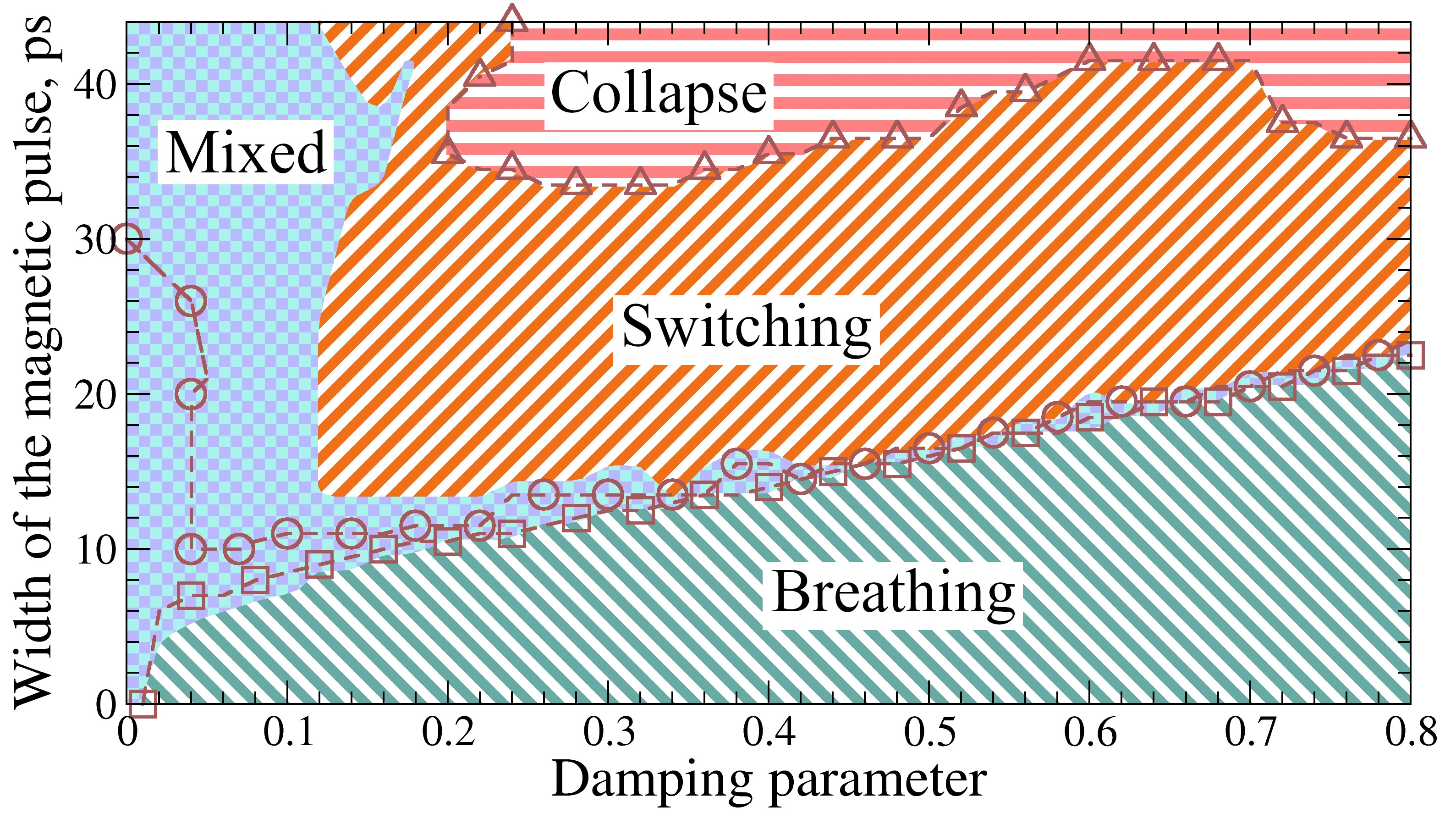} 
\caption{Process diagram obtained by using spin dynamics calculations performed with the  $\theta = 40^\circ$ magnetic pulses. Phase boundaries determined by means of neural network are indicated by brown lines.} \label{Phases40}
\end{figure}

\section{Process diagram}
\begin{figure}[t]
\center 
\includegraphics[width=\columnwidth]{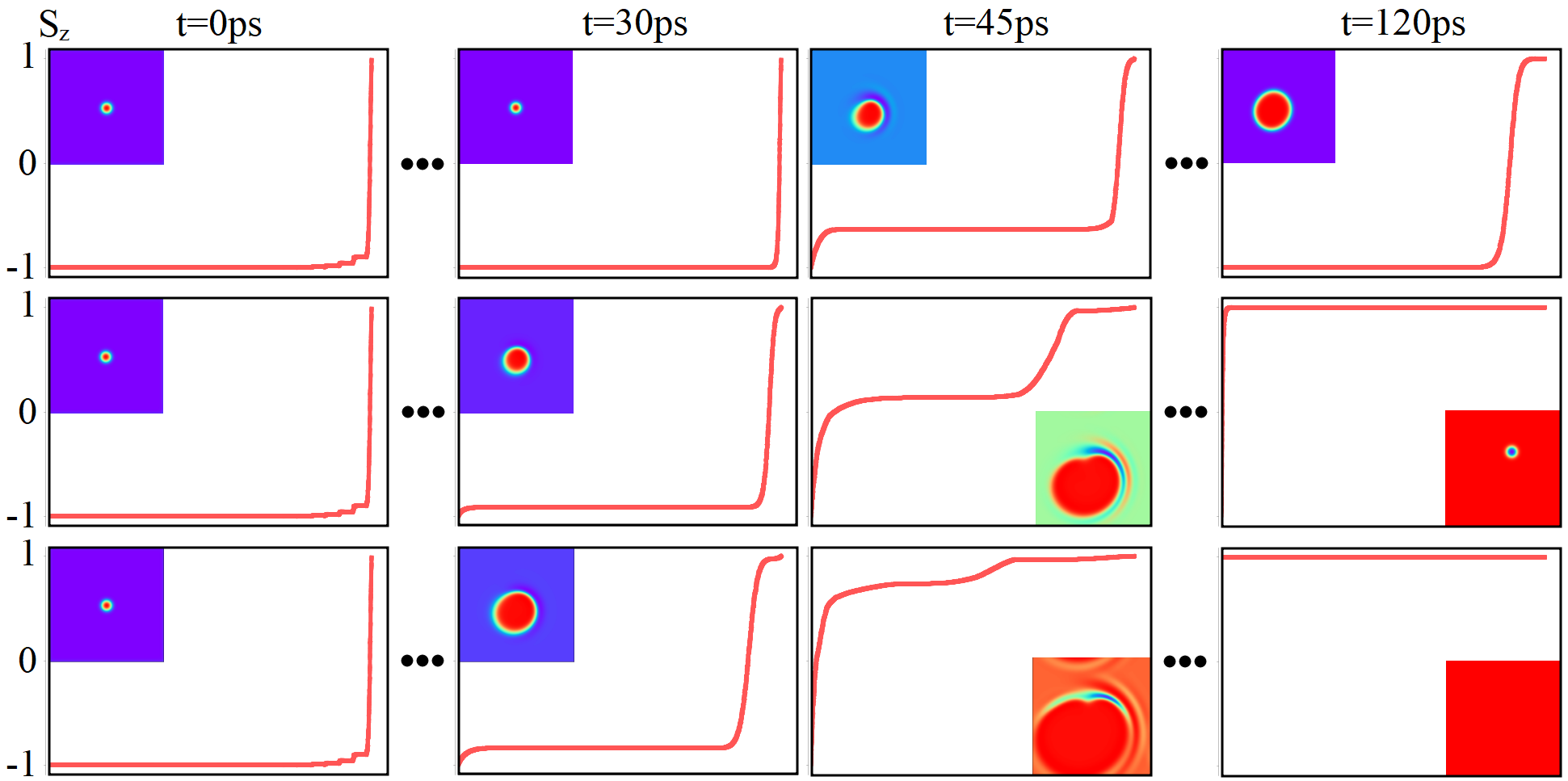} 
\caption{Examples of the different skyrmion excitation processes realized in two-dimensional systems with Dzyaloshinskii-Moriya interaction and represented as sequences of profiles. (a) breathing process generated with $\alpha$=0.36 and $t_{w}$=8 ps. (b) switching process ($\alpha$=0.36  and $t_{w}$=28 ps). (c) collapse process ($\alpha$=0.36 and $t_{w}$=40 ps). The insets are the complete magnetic configurations at specific times. The time of the maximum field pulse is 40 ps.} \label{Profiles}
\end{figure}

Fig.~\ref{Phases40} gives the switching diagram classifying the skyrmion dynamics in the system depending on the $\theta = 40^\circ$ magnetic pulse width and damping factor. To construct such a diagram we used a 44$\times$40 grid (1760 points in total) on the $t_w$/$\alpha$ plane. The process corresponding to each point of the diagram was classified manually. In accordance with the results of Ref.~\onlinecite{Blugel} there are four different types of the skyrmion switching processes: breathing, collapse, switching and mixed. Examples of the collapse, switching and breathing processes are presented in Fig.~\ref{Profiles}. The breathing phase is related to a gradual change of the skyrmion size over time. As a result the initial and final profile frames are close to each other. It is not the case for switching and collapse processes which are characterized by a global change in the system spins orientation. One can see that these processes are related to the evolution of single skyrmion in the system and clearly distinguishable on the level of the sequences of the profiles. Following Ref.~\onlinecite{Blugel} the processes for which the final state composed of multiple skyrmions and domain walls are classified as mixed state. The latter will be discussed below.

\begin{figure}[b]
\center 
\includegraphics[width=\columnwidth]{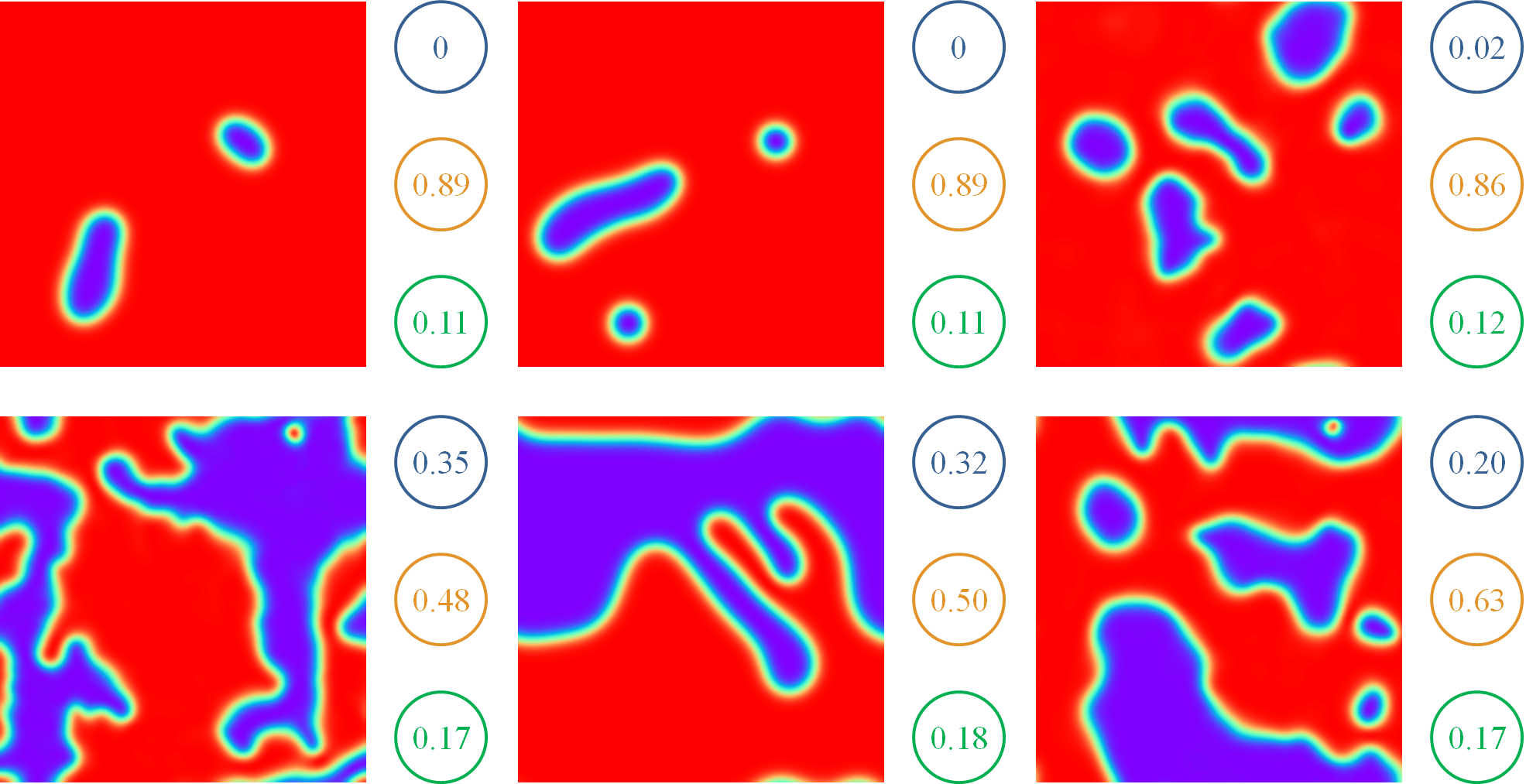} 
\caption{Examples of final magnetic configurations of the processes classified by the RNN as the switching (top panels) and mixed (bottom panels) phase. Numbers in blue, orange and green circles correspond to the values of breathing, switching and collapse outputs of the neural network, respectively. These configurations were obtained at the parameters ($\alpha$=0.1, $t_{w}$=32 ps), ($\alpha$=0.12, $t_{w}$=40 ps), ($\alpha$=0.04, $t_{w}$=14 ps), ($\alpha$=0.04, $t_{w}$=8 ps), ($\alpha$=0.58, $t_{w}$=18 ps), ($\alpha$=0.04, $t_{w}$=24 ps) from left to right. Within the classification introduced in Ref.~\onlinecite{Blugel} all the processes belong to the mixed phase.} \label{mixed_examples}
\end{figure}

{\it Neural network classification.}
The same Fig.~\ref{Phases40} contains processes classification performed by the trained recurrent neural network. One can see that the RNN provides very accurate classification of the breathing and collapse relaxation processes. On the other hand the neural network classifier shifts the boundary between mixed and switching processes to smaller values of the damping parameter. Within the manual classification discussed above a switching process will be associated with the mixed phase either if the system hosts more than one skyrmion or the final configuration is characterized by a combination of the skyrmion and domain wall. Some examples of the processes that we define as mixed ones are represented in Fig.~\ref{mixed_examples}. 

\begin{figure}[t]
\center 
\includegraphics[width=\columnwidth]{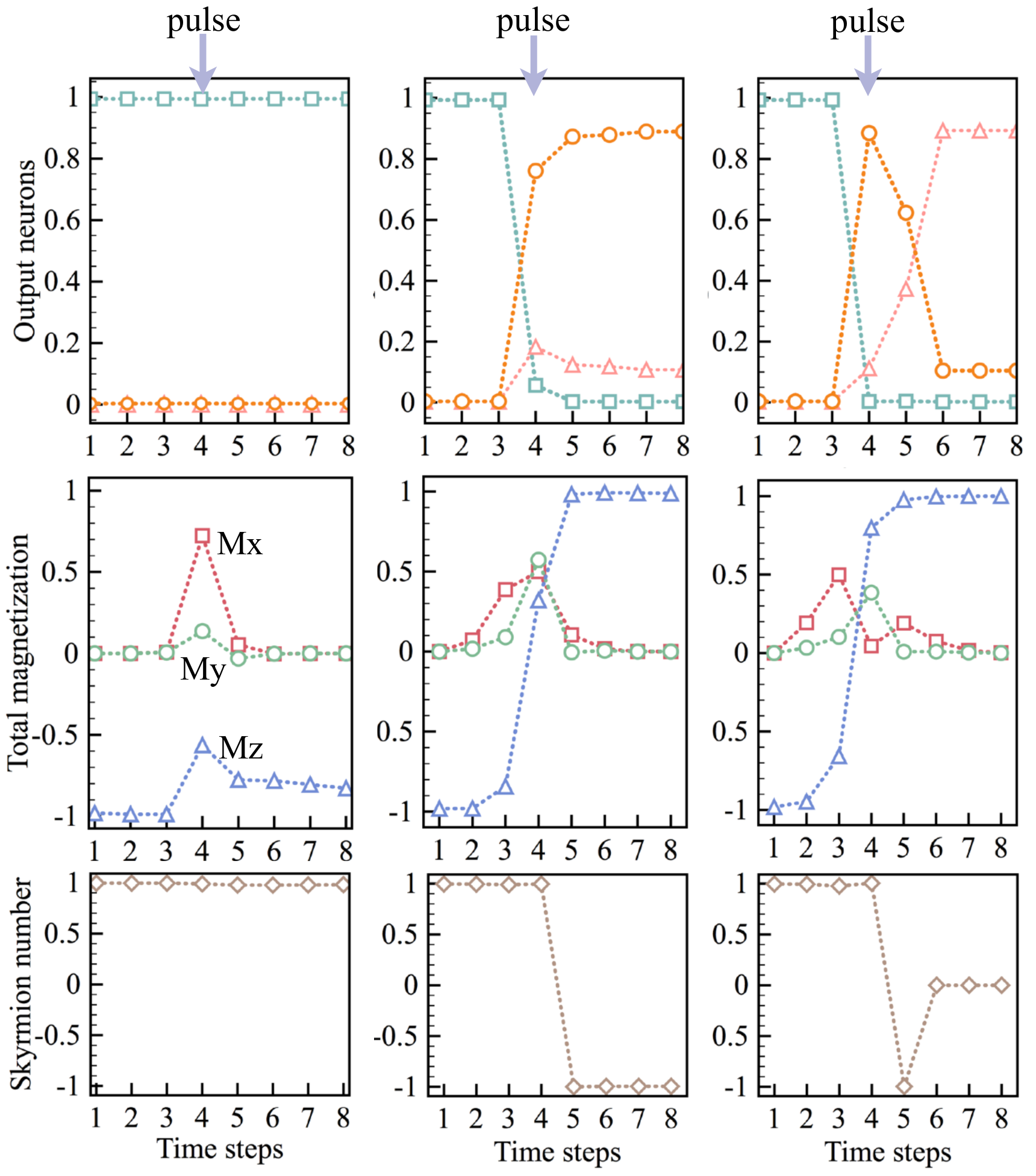} 
\caption{(Top panels) Output neurons values obtained at different time steps for breathing (left), switching (middle) and collapse (right) processes. The duration of the time step is 15 ps. The processes were generated with $\alpha$=0.36 and $t_{w}$=8 ps (breathing), $\alpha$=0.36 and $t_{w}$=28 ps (switching), $\alpha$=0.36 and $t_{w}$=40 ps (collapse). Square, circle and triangle symbols denote the neural network outputs corresponding to breathing, switching and collapse processes, respectively. (Middle panels) The components of the total magnetization calculated at each time step. Red, green and blue lines correspond to $M_x$, $M_y$ and  $M_z$, respectively. (Bottom panels) The calculated skyrmion numbers.}\label{fig:outputs}
\end{figure}

Since the neural network does not distinguish the skyrmion configurations with different topological charges (Fig.~\ref{mixed_examples} top panels), the switching processes accompanied with creation of two or more skyrmions are classified as switching ones, which is not the case for manual classification introduced in Ref.~\onlinecite{Blugel}. Thus, the neural network has refined the process diagram boundaries. However, the network is able to identify correctly configurations which are far from the skyrmion ones. They are mostly located at small damping parameter and between clear breathing and switching phases. Typical example is given in Fig.~\ref{mixed_examples} bottom panels. For such processes the values of output neurons are less than the threshold which is equal to $0.8$. 

A narrow parameters area of the process diagram between switching phase defined manually and mixed phase defined by the RNN is of technological importance since one can create two and more skyrmions from single skyrmion by magnetic pulses. Previously, the duplication mechanism of magnetic skyrmions was found and analyzed in Ref.~\onlinecite{Blugel2}. We found that there are many small regions in phase diagram Fig.~\ref{Phases40} in which it is possible to observe the creation of two skyrmions. The biggest one is located at the parameters $\alpha \in (0.09; 0.12)$ and $t_w \in (16; 32) ps$. 

\begin{figure}[t]
\center 
\includegraphics[width=\columnwidth]{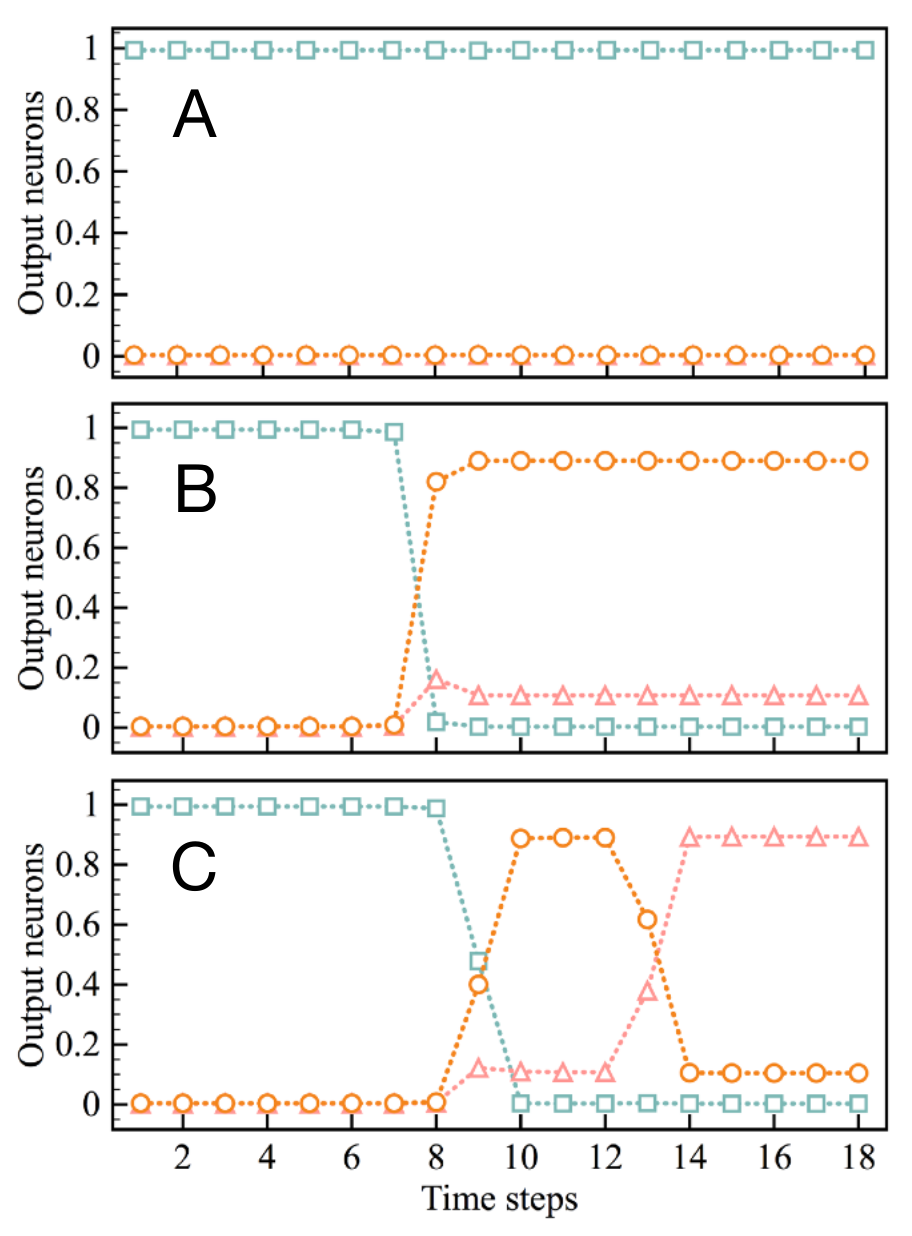} 
\caption{Demonstration of the trained neural network performance by the example of the data generated on the time scale that differ from the training one. Square, circle and triangle symbols denote the neural network outputs corresponding to breathing, switching and collapse processes, respectively. There are 18 time steps of 5 ps. (Top) The breathing process generated with $\alpha$=0.36 and $t_{w}$=8 ps. (Middle) The switching process ($\alpha$=0.36 and $t_{w}$=28 ps). (Bottom) The collapse process ($\alpha$=0.36 and $t_{w}$=40 ps).  } \label{18steps}
\end{figure}

One of the main advantages of the RNN is the ability of classification of the whole time sequence of the magnetization frames instead of individual ones. To examine whether it works in our case, we calculated the values of the output neurons at each time step and analyzed the following characteristics for different processes: output neuron values, components of the total magnetization and skyrmion number. As can be seen from Fig.~\ref{fig:outputs} (top panels) at the beginning all the configurations are classified as the breathing state because we used $softmax$ function for the output layer and the sum of all outputs has to be equal to unity. Besides, initial state has the negative magnetization which is one of the key features of the breathing process. All changes start at the fourth time step, which corresponds to 45 ps in the considered case. In other words, the RNN responds to the magnetic pulse. It is also clearly seen that the values of the output neurons change slowly step by step, which means that network relies not only on the current magnetic configuration. 

Since we used the only $z$ components of spins the question arises whether the RNN with single hidden layer performs classification on the basis of the total $M_z$. We address it by plotting the magnetization and topological charge behaviour for the same processes. One can see from Fig.~\ref{fig:outputs} (middle panels) that the switching and collapse have almost the same $M_z$ profiles while the network outputs are significantly different. Indeed, for collapse process we have a large value on the switching neuron at time steps 5 and 6 before vanishing the switched skyrmion. However, we can traced out the similar behaviour of topological charge which was not given to RNN. It gives us the opportunity to conclude that the network learns some topological parameters of the magnetic configurations.

\section{Network tests}

{\it Variation of the time scale.} Since we have trained the RNN on the processes with the fixed number of the time steps and center of the Gaussian magnetic pulse another interesting question whether we can apply it to classify  processes obtained at different parameters. The answer is yes we can! We found that it does not matter how many time steps we have in our process and when we act on the system with the magnetic pulse. To illustrate this we plotted the network outputs for the same processes we discussed in previous paragraph, but with the parameters $t = 90$ ps and $\Delta t = 5$ ps instead of $t = 105$ ps and $\Delta t = 15$ ps. Since we used the same $t_p = 40$ ps in both cases maximum of the magnetic pulse moved from the fourth to the eighth time step. As can be seen from Fig.~\ref{18steps} all the dependencies have the same behaviour as in Fig.~\ref{fig:outputs} (top panels). It gives us the opportunity to use our approach for classification of the experimental ultra-fast processes.

\begin{figure}[t]
\center 
\includegraphics[width=\columnwidth]{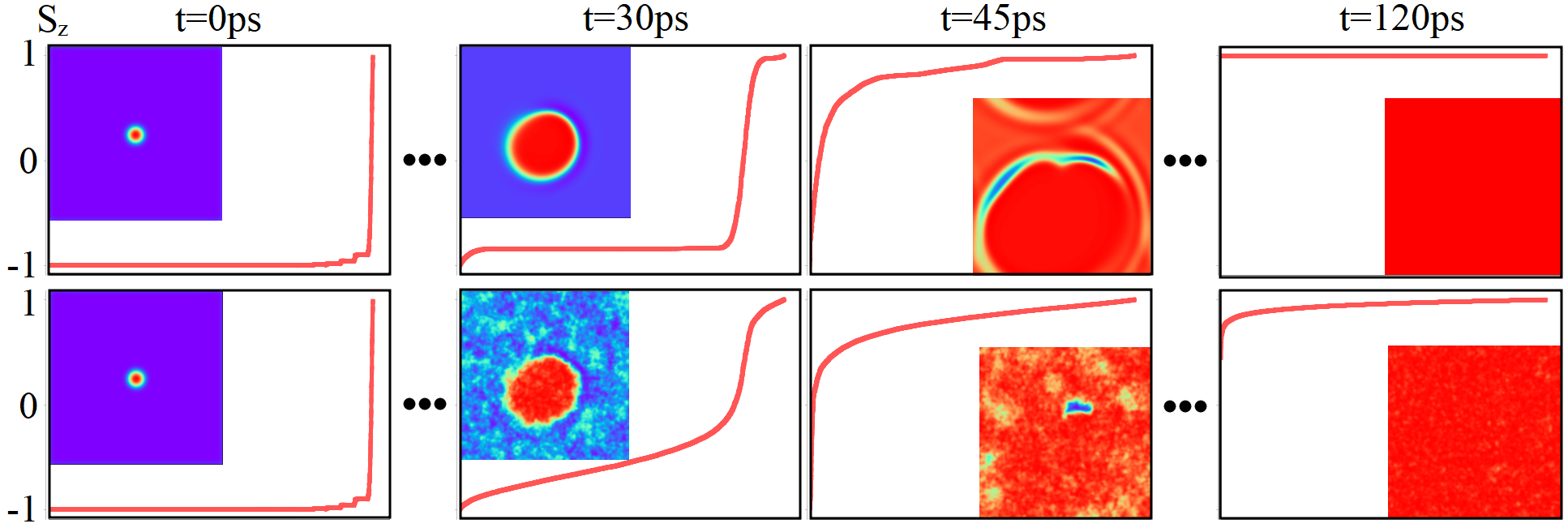} 
\caption{Comparison of the collapse processes generated at T = 0 (top) and T = 1.5 K (bottom). The parameters for these simulations were chosen to be $\alpha$=0.22 and $t_{w}$=38 ps.} \label{a22profile}
\end{figure}

\begin{figure}[b]
\center 
\includegraphics[width=\columnwidth]{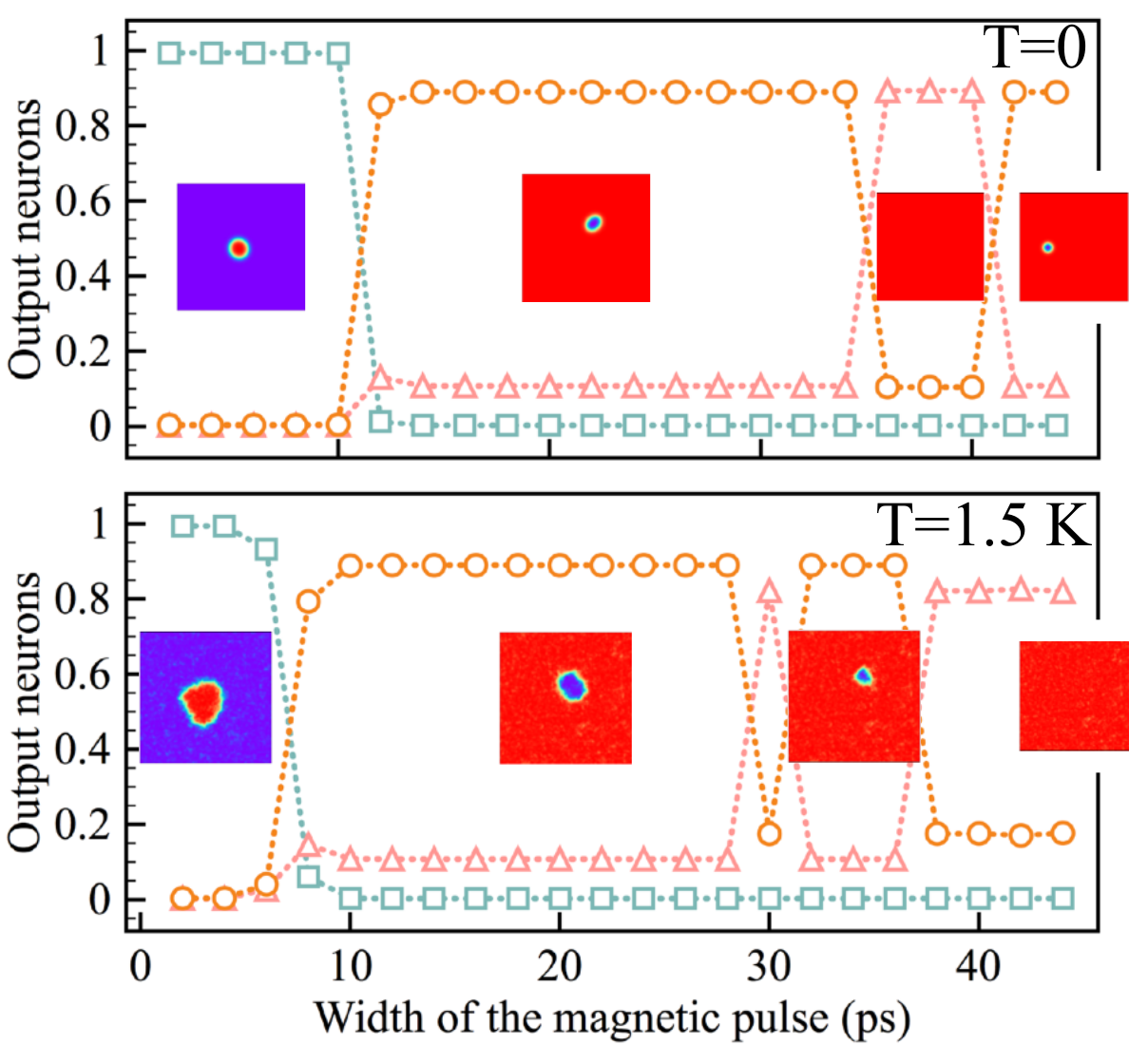} 
\caption{Comparison of the neural network outputs depending on the width of the magnetic pulse obtained at different temperatures. The insets represent the final magnetic configuration of the processes classified by the neural network. Square, circle and triangle symbols denote the neural network outputs corresponding to breathing, switching and collapse processes, respectively.} \label{temperature}
\end{figure}

{\it Temperature effect.}
The results we present above were obtained at the zero temperature. We found that the trained network can perform a robust and accurate classification of the magnetic processes up to 1.5 K. The example of this is presented in Figs.~\ref{a22profile} and \ref{temperature}.  Taking into account that the transition to the paramagnetic state is at about 6 K, we conclude that one can use the zero-temperature trained network for analysis of the low-temperature processes. It follows from Fig.~\ref{temperature} that within chosen range of magnetic pulse width values the process diagram changes its shape at the finite temperature. The transition from breathing to switching phase shifts towards smaller values of $t_w$. At high values of the pulse width ($t_w >$ 40 ps) we observe the formation of the collapse phase instead of switching one at T = 0.

\section{Conclusions} 
The neural network approach we propose is general nature and can be used for classification of different dynamical processes realized in real magnetic materials and systems. For instance, the first problem closely related to that we consider here is the refinement of the process boundaries for vortex dynamics reported in Ref.~\onlinecite{vortex}. It is also not limited by two-dimensional case, the profile procedure we used for preprocessing magnetization data gives reliable classification results in the static case for three-dimensional magnets with Dzyaloshinskii-Moriya interaction \cite{Iakovlev2}. The generalization power of the network can be increased by adding more hidden neurons layers. In this case one would expect that the network will correctly classify processes at high temperatures close to the paramagnetic phase.
\section{Acknowledgements} 
We would like to thank Yaroslav Kvashnin and Anders Bergman for fruitful discussions and technical assistance with UppASD package. This work was supported by the Russian Science Foundation Grant 18-12-00185.

\appendix

\section{Network details}

The recurrent neural network consists of three layers: input, hidden and output. The hidden layer is self cycled, which means that the states of the hidden neurons pass forward to the next steps as shown in Fig.~\ref{rnn}. There are 3 weight matrices, $V$ is for connecting input layer and hidden layer neurons, $W$ is for connecting the hidden layer and output layer neurons, and $U$ is for connecting hidden layer neurons at different time steps. Thus, the states of the hidden layer neurons depend on the current input magnetization data and previous state of itself.
The input layer is characterized by 10000 neurons, as the number of spins on the lattice. For the hidden layer we have found out that the minimal number of the neurons is 512. If the hidden layer contains less neurons, RNN is not able to reach acceptable error rate during the training procedure. The output layer has 3 neurons, by the number of the processes we want to classify.

\section{Machine learning details} 
\begin{figure}[h!]
\center 
\includegraphics[width=\columnwidth]{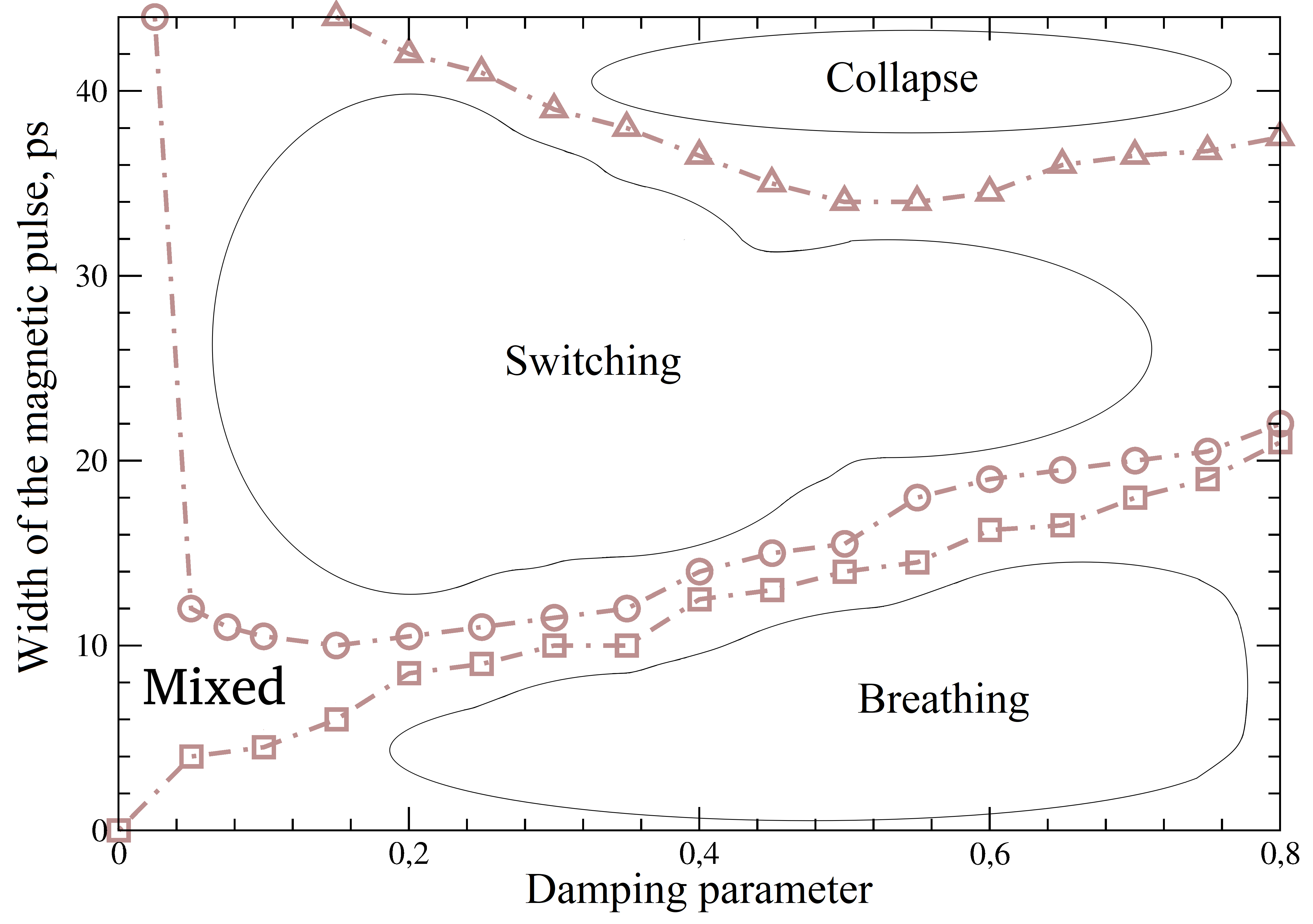} 
\caption{Process diagram constructed on the basis of the LLG simulations with the $\theta = 50^\circ$ magnetic pulses. Selected areas correspond to the processes we used for supervised learning of the recurrent neural network.} \label{Phases50}
\end{figure}

For the training set we generated 400 configurations for each of breathing, switching and collapse processes from the internal areas of the corresponding diagram Fig.~\ref{Phases50}. Such a small amount of data is due to the fact that we used the LLG equation~\eqref{LLG} to simulate the magnetization dynamics. Results of such simulations does not changes from time to time at fixed parameters in contradistinction to the Monte Carlo ones.

\begin{figure}[t]
\center 
\includegraphics[width=\columnwidth]{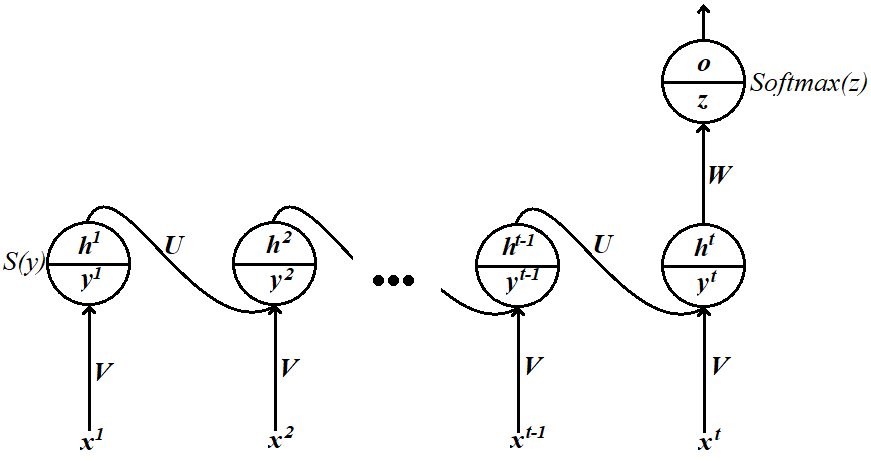} 
\caption{Scheme of recurrent neural network. All the notations are described in the text.} \label{rnn}
\end{figure}

As an input for our recurrent neural network (Fig.~\ref{rnn}) we used the sorted vectors containing $S^z$ components of a two-dimensional magnetic configurations. The hidden layer neurons activate by means of the sigmoid function,
\begin{eqnarray}
h_j^t = \sigma (y_j^t) = \frac{1}{1+e^{-y_j^t}},
\end{eqnarray}
\begin{equation}\label{hi}
y_j^t=\sum_{k=1}^N x^{t}_{k}V_{kj}+\sum_{m=1}^{N_h}h^{t-1}_{m}U_{mj},
\end{equation}
where $x_k = S_i^z$ is the value of $i$th input neuron, $V_{kj}$ is the weight between the $k$th input neuron and $j$th hidden neuron, $U_{mj}$ --- weight between the $m$th hidden neuron at time step $t-1$ and $j$th hidden neuron at time step $t$, $N=L\times L$ --- number of the input neurons, $N_h$ is the number of the hidden neurons.

In turn, for output neurons we use the softmax function that is given by 
\begin{eqnarray}
o_i = softmax(z_i) = \frac{e^{z_i}}{\sum_{n = 1}^{N_o}e^{z_n}},
\end{eqnarray}
\begin{equation}
z_{i}=\sum_{j=1}^{N_h}h^{t}_{j}W_{ji},
\end{equation}
where $N_o$ is the number of the output neurons, $W_{ji}$ --- weight between the $j$th hidden neuron and $i$th output neuron.

During the learning process, we randomly choose 10\% of training set for cross-validation to avoid overfitting and define the stopping point where error is less than the required value. To evaluate the error we used cross entropy function given by the following expression:
\begin{equation}\label{Error}
E = -\sum_{i=1}^{N_o}t_i\log o_i,
\end{equation}
where $t_i$ represents the ground truth labels.

The network optimization was fulfilled through back-propagation through time algorithm \cite{bptt}. All weight matrices were initialized by means of the ratio of 1 divided by a number of neurons in the previous layer to avoid extremely large sigmoid arguments: $V_{kj}=\frac{1}{10000}$, $W_{ji}=\frac{1}{512}$, $U_{mj}=\frac{1}{512}$.  We used the following expressions for new weights in order to avoid getting stuck in a local minima 

\begin{equation}\label{dw}
W_{ji}=W_{ji}+\frac{\delta E}{\delta W_{ij}},
\end{equation}
\begin{equation}\label{dv}
V_{kj}=V_{kj}+\sum_{p=0}^{N_t-1}\frac{\delta L^{t-p}_{j}}{\delta V_{kj}},
\end{equation}
\begin{equation}\label{du}
U_{mj}=U_{mj}+\sum_{p=0}^{N_t-1}\frac{\delta L^{t-p}_{j}}{\delta U_{jm}},
\end{equation}
where $N_t$ is the total number of time steps in our process, $p$ shows the index of the previous time step with respect to the final one. In these expressions $\frac{\delta E}{\delta W}$, $L_{j}^{t-p}$, $\frac{\delta L}{\delta V}$ and $\frac{\delta L}{\delta U}$ are given by the equations below. All the additional indices here we introduced to show the summation rules. For simplicity the equations for $\delta L^{t-p}_{n}$ were written in recurrent form.

\begin{equation}\label{dedw}
\frac{\delta E_{i}}{\delta W_{ij}}=(t_{i}-o_{i})h_{j},
\end{equation}

\begin{equation}\label{Lt}
L_{n}^{t-p} = \underbrace{\sum_{j,m,l,...,n=1}^{N_h}}_{p}\sum_{i=1}^{N_o}E_iW_{ij}\underbrace{(U_{jm} U_{ml} ... U_{cn})}_{p},
\end{equation}

\begin{equation}\label{dldv}
\frac{\delta L^{t}_{j}}{\delta V_{kj}}=\sum_{i}(W_{ji}^{2}\frac{\delta E_{i}}{\delta W_{ij}})(1-h_{j}^{t})x_{k}^{t},
\end{equation}

\begin{equation}\label{dldu}
\frac{\delta L^{t}_{j}}{\delta U_{mj}}=\sum_{i}(W_{ji}^{2}\frac{\delta E_{i}}{\delta W_{ij}})(1-h_{j}^{t})h_{k}^{t-1},
\end{equation}

\begin{equation}\label{dltpdv}
\frac{\delta L^{t-p}_{n}}{\delta V_{kn}}=\sum_{q}^{N}\sum_{r}^{N_{h}}(U_{nr}^{2} \frac{\delta L^{t-p+1}_{r}}{\delta V_{qr}} x^{t-p+1}_{q})h_{n}^{t-p}(1-h_{n}^{t-p})x_{k}^{t-p},
\end{equation}

\begin{equation}\label{dltpdu}
\frac{\delta L^{t-p}_{n}}{\delta U_{ln}}=\sum_{q,r}^{N_{h}}(U_{nr}^{2} \frac{\delta L^{t-p+1}_{r}}{\delta U_{qr}} h^{t-p+1}_{q})h_{n}^{t-p}(1-h_{n}^{t-p})h_{l}^{t-p-1}.
\end{equation}

\end{document}